\documentclass[
preprint,
  nofootinbib,amsmath,amssymb,aps,preprintnumbers,superscriptaddress,notitlepage,
]{revtex4-2}

\usepackage{graphicx}
\usepackage{bm}
\usepackage{braket}
\usepackage{cases} 
\usepackage{here}
\usepackage{color}
\usepackage{upgreek}
\usepackage{ulem}
\usepackage{comment}
\usepackage{empheq}
\usepackage{bm}

\begin{document}
\begin{flushright}
KEK-QUP-2022-0018, KEK-TH-2477, KEK-Cosmo-0304, KOBE-COSMO-22-19
\end{flushright}

\title{Exploring High Frequency Gravitational Waves with Magnons}

\author{Asuka Ito}
\email[]{asuka.ito@kek.jp}
\affiliation{International Center for Quantum-field Measurement Systems for Studies of the Universe and Particles (QUP), KEK, Tsukuba 305-0801, Japan}
\affiliation{Theory Center, Institute of Particle and Nuclear Studies, KEK, Tsukuba 305-0801, Japan}
\author{Jiro Soda}
\email[]{jiro@phys.sci.kobe-u.ac.jp}
\affiliation{Department of Physics, Kobe University,
 Kobe 657-8501, Japan}

\vspace{2.0cm}

\date{\today}

\begin{abstract}
Detecting gravitational waves with frequencies higher than 10 kHz requires new strategies.
In previous papers, we proposed magnon gravitational wave detectors and gave the first limit on GHz 
gravitational waves by reinterpreting the existing data from
axion dark matter experiments. 
In this paper, we show that the sensitivity can be improved by constructing the detector specific to gravitational waves.
In particular,  we employ an infinite sum of terms in the expansion of Fermi normal coordinates to probe gravitational waves with a wavelength comparable to the detector
size.  As a consequence, we obtain sensitivity of around $h_c \sim 10^{-20}$.
\end{abstract}

\maketitle

\newpage \tableofcontents
%
%
%
%
%
%
%

\section{Introduction}

The first direct detection of gravitational waves (GWs) has opened up the era of GW astrophysics~\cite{LIGOScientific:2016aoc}.
Ground interferometers can probe a frequency range from 1Hz to 1kHz. 
Future space GW interferometers will boost GW astrophysics and promote the development of
GW cosmology. Space detectors extend the observable frequency to a low frequency range up to $10^{-4}$ Hz,
and pulsar timing arrays can probe the nHz range.  Thus, there has been substantial development in low frequency observations. 
However, since we can expect that a new frequency will give rise to a new discovery, 
it is important to extend the observable frequency range not only to the low frequency range but also to the high frequency range.

From a theoretical perspective,  we need high frequency GWs to explore fundamental physics~\cite{Aggarwal:2020olq}.
There are three categories for GW sources depending on the mechanism of generation, namely, inflationary, cosmological, and astrophysical sources.
The frequency of primordial GWs generated during inflation would range from $10^{-18}$ Hz to $1$ GHz.
After inflation, there are many cosmological sources containing information on particle physics. 
These phenomena in the early universe can produce GWs below GHz frequencies. 
For example, 
we know that the first order phase transition around $10^{2}$ GeV
can generate GWs with frequency around $10^{-5}$ Hz~\cite{Ellis:2018mja}. 
Simple scaling argument tells us that the first order phase transition at an energy scale below the GUT scale
can generate GWs with frequencies below the GHz range.
Finally, 
GWs emitted from astrophysical sources are determined by the mass of the object.  
Strong GWs can be expected from massive compact objects such as black holes.
It is believed that the minimum mass of astrophysical objects is around the solar mass. Hence, the maximal frequency of astrophysical GWs would be $10$ kHz.
However, there may exist primordial black holes whose mass is less than the solar mass. 
Intriguingly, primordial black holes with a mass below the solar mass can be probed through the observation of high 
frequency GWs~\cite{Franciolini:2022htd}.
It is also argued that GHz GWs can be used to probe the end of inflation or to rule out inflation~\cite{Ito:2020neq,Vagnozzi:2022qmc}.

It should be noted that the experimental constraints on
axions can be reinterpreted as constraints on high frequency GWs. 
Indeed, in the GHz range, we have imposed constraints on GHz GWs 
by utilizing the existing data for axion searches with magnons~\cite{Ito:2019wcb,Ito:2020wxi}. 
Subsequently, this strategy has been extended to other axion search experiments which utilize
phenomena due to the axion-photon interaction, such as the axion-photon conversion~\cite{Ejlli:2019bqj,Domcke:2022rgu} and
the resonant excitation of photons by the axion dark matter~\cite{Berlin:2021txa}.

As we mentioned above, the same system can potentially detect both axions and GWs potentially.
In other words, they can each be contamination to the other.
Thus, a natural question is whether we can distinguish the two effects.
It should be noted that axion cavity experiments use axion-photon coupling, 
while axion magnon experiments use the axion-electron coupling.
On the other hand, GWs universally affect both photons and electrons. 
Therefore, if information regarding both is available, it is possible to discriminate between axions and gravitons in principle.
Thus, the sensitivity of magnon GW detectors is worth studying.

This paper is organized as follows.
In section II, we review Fermi normal coordinates in a planar GW.
In section III, we explain how the planar GW excites magnons.
We estimate the sensitivity of magnon GW detectors in section IV.
The final section is devoted to the conclusion.

\section{Fermi normal coordinates in a planar GW} \label{ISS}
In order to investigate response of magnons to a planar GW, 
we need to use coordinates which comove with the magnon system.
To this end, we used Fermi normal coordinates in our previous work~\cite{Ito:2020wxi}.
In~\cite{Ito:2020wxi}, we truncated the infinite series in Fermi normal coordinates 
at the leading order by demanding that the typical size of the magnon system $l$ is much smaller than the wavelength of 
GWs $\lambda$.
However, the approximation is not accurate when $l$ is close to $\lambda$, and not valid when $l \gg \lambda$.
In such situations, one needs to take into account the infinite number of terms in the series in the Fermi normal 
coordinates~\cite{Fortini198237,Marzlin:1994ia,Rakhmanov:2014noa}.
As was shown in~\cite{Licht:2004ix,Berlin:2021txa}, the sum of the infinite series can be cast into a closed analytic form for a 
planar GW.
Actually, when one considers a planar GW  
$h_{ij} \propto \cos(\omega t - \bm{k}\cdot\bm{x})$ in the  transverse traceless gauge,
the corresponding metric in the Fermi normal coordinate is given by
%
%
\begin{equation}
  \begin{cases}
    g_{00} = -1 - 2 R_{0i0j}|_{\bm{x}=0} \, x^{i}x^{j} \times {\rm Re}
            \Big[ \frac{1 - e^{-i\bm{k}\cdot\bm{x} }}{\left( \bm{k}\cdot\bm{x} \right)^{2} } 
            -  \frac{i}{(\bm{k}\cdot\bm{x})} \Big] ,
    &  \\
    g_{0i} = 2 R_{0jik}|_{\bm{x}=0} \, x^{j}x^{k} \times {\rm Re}
            \Big[ i \frac{1 - e^{-i\bm{k}\cdot\bm{x} }}{\left( \bm{k}\cdot\bm{x} \right)^{3} } 
            +  \frac{i}{2 (\bm{k}\cdot\bm{x})} 
            +  \frac{e^{-i\bm{k}\cdot\bm{x}}}{\left( \bm{k}\cdot\bm{x} \right)^{2}}     \Big] ,
    &  \\
    g_{ij} = \delta_{ij} + 2 R_{ikjl}|_{\bm{x}=0} \, x^{k}x^{l} \times {\rm Re}
            \Big[ \frac{1 + e^{-i\bm{k}\cdot\bm{x} }}{\left( \bm{k}\cdot\bm{x} \right)^{2} } 
            + 2i \frac{1 - e^{-i\bm{k}\cdot\bm{x}}}{\left( \bm{k}\cdot\bm{x} \right)^{3}} \Big] ,
    &
      \end{cases}
      \label{met}
\end{equation}
where the Riemann tensor is defined by
\begin{equation}
    R^{\alpha}{}_{\mu\beta\nu} 
             =  \frac{1}{2}(h^{\alpha}_{\ \nu,\mu\beta}-h_{\mu\nu\ ,\beta}^{\ \ ,\alpha}
                   -h^{\alpha}_{\ \beta,\mu\nu}+h_{\mu\beta\ ,\nu}^{\ \ ,\alpha}) \ , \label{Rie2}
\end{equation}
and evaluated at the origin of the coordinate.

We use the Fermi normal coordinate with the metric (\ref{met}) to formulate excitations of magnons induced by a planar GW, which 
is applicable to any magnitude of $|\bm{k}\cdot\bm{x}| \sim kl$, even for the value larger than unity.
Note that, in practice, we will use the transverse-traceless gauge to calculate the Riemann tensor since 
it is gauge invariant at linear order.
In previous papers~\cite{Ito:2020wxi,Ito:2020xvp}, interaction between a spin of an electron and GWs in the presence of an external magnetic 
field was derived by taking the non-relativistic limit of the Dirac equation.
The Hamiltonian for the interaction is given by
\begin{equation}
  H_{{\rm GW}} =
  - \mu_{B} B^{i}  \hat{S}^{j}  Q_{ij} \ ,  \label{qab}
\end{equation}  
where $\mu_{B}$ is the Bohr magneton, and $\hat{S}$ stands for a spin operator of a Dirac particle such as an electron.
The effect of a planar GW is represented by $Q_{ij}$.
The explicit formula derived in~\cite{Ito:2020wxi} is 
  the leading approximation in the series of $|\bm{k}\cdot\bm{x}|$ in
Eq.\,(\ref{met}).
For the full metric (\ref{met}), one can repeat the discussion of~\cite{Ito:2020wxi} and obtain 
\begin{eqnarray}
  Q_{ij} &=& 2 \delta_{ij} R_{0k0l}|_{\bm{x}=0} x^{k}x^{l} \times {\rm Re}
            \Big[ \frac{1 - e^{-i\bm{k}\cdot\bm{x} }}{\left( \bm{k}\cdot\bm{x} \right)^{2} } 
         -  \frac{i}{(\bm{k}\cdot\bm{x})} \Big]  \nonumber \\
         && -2 \Big[ \delta_{ij} R_{mkml}|_{\bm{x}=0} x^{k}x^{l} - R_{ikjl}|_{\bm{x}=0}  x^{k}x^{l} \Big]
            \times {\rm Re}
            \Big[ \frac{1 + e^{-i\bm{k}\cdot\bm{x} }}{\left( \bm{k}\cdot\bm{x} \right)^{2} } 
            + 2i \frac{1 - e^{-i\bm{k}\cdot\bm{x}}}{\left( \bm{k}\cdot\bm{x} \right)^{3}} \Big] .  \label{qab2}
\end{eqnarray}
If one expands the parts, taking the real part in $| \bm{k}\cdot\bm{x} |$, the leading terms coincide 
with the result of~\cite{Ito:2020wxi}.
In the next section, we will investigate resonant excitations of magnons by a planar GW by utilizing Eqs.\,(\ref{qab}) and (\ref{qab2}).
\section{Magnon excitations induced by a planar GW} \label{GMreso}
In this section, we study the effect of a planar GW on magnons, which are collective spin excitations.
As an experimental setup, for example, we consider a spherical ferromagnetic crystal in an external magnetic field.
It is known that such a system is well described by the Heisenberg Hamiltonian~\cite{Heisenberg1926}.
Then, as we saw in the previous section, a planar GW can interact with each spin inside the ferromagnetic crystal 
as described by Eqs.\,(\ref{qab}) and (\ref{qab2}).
The total Hamiltonian we consider is
\begin{eqnarray}
  H_{{\rm tot}}  
       =
       - \mu_{B} \left( 2 \delta_{za} + Q_{za} \right) 
          B_{z} \sum_{i}   \hat{S}^{a}_{(i)} 
       -  \sum_{i,j} J_{ij} \hat{\bm{S}}_{(i)} \cdot \hat{\bm{S}}_{(j)} \ ,  \label{total}
\end{eqnarray}
where we applied an external magnetic field along the $z$-direction, $B_{z}$, and
$\hat{S}^{a}_{(i)}$ represents the spin operators of electrons on each site specified by $i$. 
The first term is the conventional Pauli term, which turns the spin direction to that of
the external magnetic field.
The third term represents the exchange interactions
between spins with strength $J_{ij}$.
The second term describes the effect of a planar GW
on a spin located at $x^{i}$ in the Fermi normal coordinate whose origin comoves with the center of mass of the crystal.
Indeed, at the origin, $x^{i}=0$, we see that $Q_{ij}=0$ conforms to the equivalence principle.

The spin system (\ref{total}) can be rewritten by using
the Holstein-Primakoff transformation~\cite{Holstein:1940zp}: 
\begin{eqnarray}
  \begin{cases}
   \hat{S}_{(i)}^{z} = \frac{1}{2} - \hat{C}_{i}^{\dagger} \hat{C}_{i} \ , & \\
   \hat{S}_{(i)}^{+} = \sqrt{1- \hat{C}_{i}^{\dagger} \hat{C}_{i}} \ \hat{C}_{i}  \ , & \\
   \hat{S}_{(i)}^{-} = \hat{C}_{i}^{\dagger} \sqrt{1- \hat{C}_{i}^{\dagger} \hat{C}_{i}} \ , &
  \end{cases} \label{prim}
\end{eqnarray}
where  $\hat{C}_{i}$ and $\hat{C}^{\dagger}_{i}$ are bosonic annihilation and creation operators and
$\hat{S}_{(j)}^{\pm} = \hat{S}_{(j)}^{x} \pm i \hat{S}_{(j)}^{y}$ are the ladder operators.
Actually, one can confirm that the SU(2) algebra, $[\hat{S}^{i} , \hat{S}^{j}] = i \epsilon_{ijk} \hat{S}^{k}$ 
($i,j,k = x, y, z$), using the commutation relations $[\hat{C}_{i}, \hat{C}^{\dagger}_{j}]=\delta_{ij}$.
Furthermore, provided that contributions from the surface of the crystal are negligible,
one can expand the bosonic operators by plane waves as 
\begin{equation}
  \hat{C}_{i} = \sum_{\bm{k}} \frac{e^{-i\bm{k} \cdot \bm{r}_{i}}}{\sqrt{N}} \hat{c}_{k} \ , \label{bbb}
\end{equation}
where $\bm{r}_{i}$ is the position vector of the $i$ spin and $N$ is the number of the spins.
An excitation of the spin waves corresponds to a particle created by $\hat{c}_{k}^{\dagger}$.
Such a quasi particle is called a magnon.

We now rewrite the spin system (\ref{total})
by magnons with the Holstein-Primakoff transformation (\ref{prim}), and 
then we focus only on the homogeneous mode of magnons, i.e., $k = 0$ mode. 
Then, the third term in the total Hamiltonian (\ref{total}) is irrelevant because
it does not contribute to the homogeneous mode.
Furthermore, because $Q_{zz}$ does not contribute to the resonance of the spins, namely the excitation of magnons,
we will drop it. 
Thus, we have
\begin{equation}
  H_{{\rm tot}} =      
        \mu_{B}  B_{z} \sum_{i}  
      \left[  2 \hat{C}_{i}^{\dagger} \hat{C}_{i}  + 
      \frac{\hat{C}_{i} + \hat{C}_{i}^{\dagger}}{2}  Q_{zx}
    + \frac{\hat{C}_{i} - \hat{C}_{i}^{\dagger}}{2i} Q_{zy} \right]  \ . 
    \label{toHami}
\end{equation}

Now let us consider a planar GW propagating in the $z$-$x$ plane so that
the wave number vector of the GW $\bm{k}$ has a direction 
$\hat{k} = ( \sin\theta , 0 , \cos\theta )$.
We can expand the GW $h_{ij}$ in terms of linear polarization tensors satisfying 
$e^{(\sigma)}_{ij} e^{(\sigma')}_{ij} = \delta_{\sigma\sigma'}$ as
\begin{equation}
  h_{ij}(\bm{x}, t) = h^{(+)}(\bm{x}, t) e^{(+)}_{ij} 
           + h^{(\times)}(\bm{x}, t) e^{(\times)}_{ij} \ . \label{expan}
\end{equation}
More explicitly, we took the representation 
\begin{numcases}
  {}
  h^{(+)}(\bm{x}, t) = \frac{h^{(+)}}{2} 
      \left( e^{-i(w_{h}t - \bm{k}\cdot\bm{x})} 
           + e^{i(w_{h}t - \bm{k}\cdot\bm{x})} \right) \ , & \\
  h^{(\times)}(\bm{x}, t) = \frac{h^{(\times)}}{2} 
        \left( e^{-i(w_{h}t - \bm{k}\cdot\bm{x} + \alpha)} 
             + e^{i(w_{h}t - \bm{k}\cdot\bm{x} +\alpha)} \right) \ , \label{cross} & 
\end{numcases}
where $\omega_{h}$ is the angular frequency of the GW, and $\alpha$ represents 
the difference of the phases of the polarization.
Note that the polarization tensors  can be explicitly constructed as
\begin{eqnarray}
  e _{ij}^{(+)} &=& \frac{1}{\sqrt{2}}\left(
    \begin{array}{ccc}
      \cos^{2}\theta & 0 & -\cos\theta \sin\theta \\
      0 & -1 & 0 \\
      -\cos\theta \sin\theta & 0 & \sin^{2}\theta
    \end{array} 
  \right) , \label{lipo1}  \\
  e _{ij}^{(\times)} &=& \frac{1}{\sqrt{2}}\left(
    \begin{array}{ccc}
      0 & \cos\theta & 0 \\
      \cos\theta & 0 & -\sin\theta \\
      0 & -\sin\theta & 0
    \end{array} 
  \right) .  \label{lipo2}
\end{eqnarray}
In the above Eqs.\,(\ref{lipo1}) and (\ref{lipo2}), we defined the $+$ mode as a deformation in the $y$-direction.

\begin{figure}[ht]
\centering
\includegraphics[width=7.5cm]{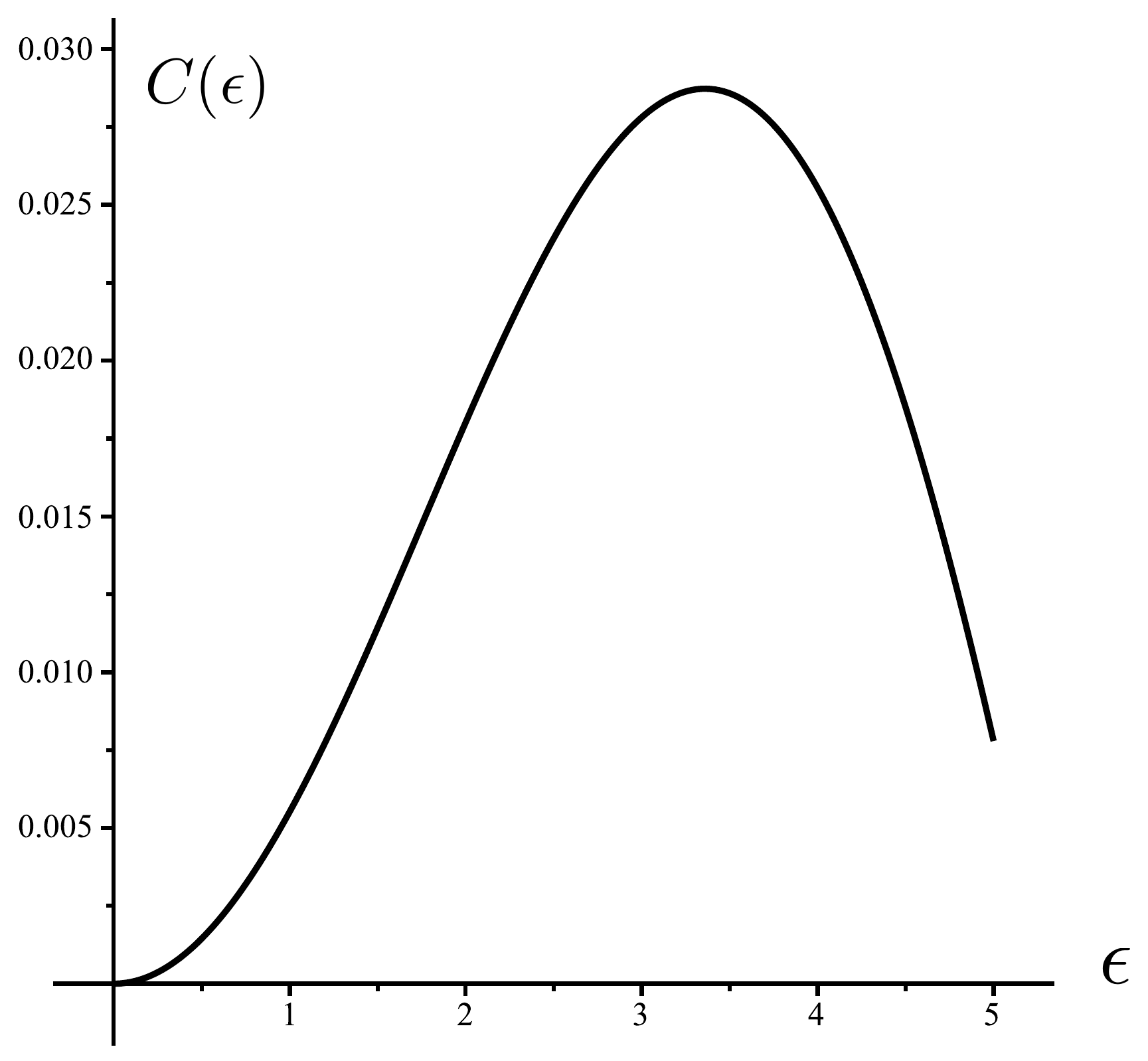}
\caption{$\epsilon = k\ell =\frac{l}{\lambda/2\pi}$ dependence of the form factor $C(\epsilon)$ is illustrated. } \label{Csekibun}
\end{figure}

Then, substituting Eqs.\,(\ref{expan})-(\ref{lipo2}) into 
the total Hamiltonian (\ref{toHami}),
moving on to the Fourier space, and using the rotating wave approximation,
one can deduce
\begin{equation}
   H_{{\rm tot}} \simeq 2\mu_{B} B_{z}  \hat{c}^{\dagger} \hat{c} +  
 g_{eff} \left(    \hat{c}^{\dagger} e^{-i\omega_{h}t} +    \hat{c} e^{i\omega_{h}t} \right)  ,
                      \label{dri}
\end{equation}
where $\hat{c} = \hat{c}_{k=0}$%
\footnote{Although considering magnon modes other than the uniform mode would be interesting, 
we focus only on the uniform mode, which is the main target in most experiments.}
and
\begin{equation}
g_{eff} =  C(l,\lambda) \mu_{B} B_{z} \sin\theta \sqrt{N} 
\left[ \cos^{2} \theta \, (h^{(+)})^{2} + (h^{(\times)})^{2} + 2 \cos\theta\sin\alpha \, h^{(+)}h^{(\times)} 
\right]^{1/2}  \ , \label{effco}
\end{equation}
is the effective coupling constant between the GW and the magnons.
$C(l,\lambda)$ is the form factor determined by the spatial integration (the sum with respect to $i$), which depends
on the radius of the crystal $l$ and the wavelength of the GW $\lambda = 2\pi / \omega_{h}$.
Although the form factor can also depend on the lattice structure, we have assumed a simple cubic lattice structure for simplicity.  
More specifically, it is given by the formula
\begin{eqnarray}
  C(l,\lambda) &=& 
-\frac{3}{16\sqrt{2}\pi l^3 } \int d^3x   \left[  1+ \cos kx 
 -\frac{2 \sin k x  }{kx } \right]  \nonumber \\
   &=&-\frac{\sqrt{2}}{8} + \frac{3\sqrt{2} {\rm Si}(\epsilon)}{8\epsilon} + \frac{3\sqrt{2} \cos(\epsilon)}{4\epsilon^{2}}
   - \frac{3\sqrt{2} \sin(\epsilon)}{4\epsilon^{3}}
                  \ ,  \label{main_form}
\end{eqnarray}
where $\epsilon \equiv kl =\frac{l}{\lambda/2\pi}$.
One can find that the form factor is actually a function of only $\epsilon$, i.e., 
$C(l,\lambda) = C(\epsilon)$.
Fig.\,\ref{Csekibun} illustrates the $\epsilon$ dependence of $C(\epsilon)$.
We see that it has a peak around $\epsilon \sim \pi$, so that the form factor is maximized 
when the wavelength of the GW $\lambda$ is comparable to the diameter of the ferromagnetic crystal $2l$.
It may also be useful to give the series of $C(\epsilon)$ with respect to $\epsilon$ $(\leq 1)$, that is
\begin{equation}
  C(\epsilon) =   \frac{\sqrt{2}}{240} \epsilon^{2}
                 - \frac{3\sqrt{2}}{11200} \epsilon^{4}
                 + \frac{\sqrt{2}}{169344}  \epsilon^{6}
                 - \frac{\sqrt{2}}{13685760} \epsilon^{8} + \cdots \ .  \label{form}
\end{equation}
The first term coincides with that  derived previously in~\cite{Ito:2020wxi}.

It is worth rewriting $g_{eff}$ in terms of the Stokes parameters to obtain a coordinate-independent expression:
\begin{equation}
g_{eff} = C(\epsilon) 
          \mu_{B} B_{z} \sin\theta \sqrt{N} 
          \left[ \frac{1+\cos^{2}\theta}{2} \, I - 
          \frac{\sin^{2}\theta}{2} \, Q  + \cos \theta \, V  \right]^{1/2}  \ , \label{effco2}
\end{equation}
where the Stokes parameters are defined~\cite{Seto:2008sr}, similarly to the electromagnetic case, by 
\begin{eqnarray}
  \begin{cases}
   I = (h^{(+)})^{2} + (h^{(\times)})^{2} \ , & \\
   Q = (h^{(+)})^{2} - (h^{(\times)})^{2}  \ , & \\
   U = 2 \cos\alpha \, h^{(+)} h^{(\times)}  \ , & \\
   V = 2 \sin\alpha \, h^{(+)} h^{(\times)}  \ . &
  \end{cases} \label{stokes}
\end{eqnarray}
They satisfy $I^{2} = U^{2} + Q^{2} +V^{2}$.
We see that the effective coupling constant depends on the polarization.
Note that the stokes parameters $Q$ and $U$ transform as 
\begin{equation}
\binom{Q'}{U'} = \begin{pmatrix}
\cos 4\psi & \sin 4\psi \\
-\sin 4\psi & \cos 4\psi   \end{pmatrix}
\binom{Q}{U}
\end{equation}
where $\psi$ is the rotation angle around $\bm{k}$.

\begin{figure}[H]
\centering
\includegraphics[width=7.5cm]{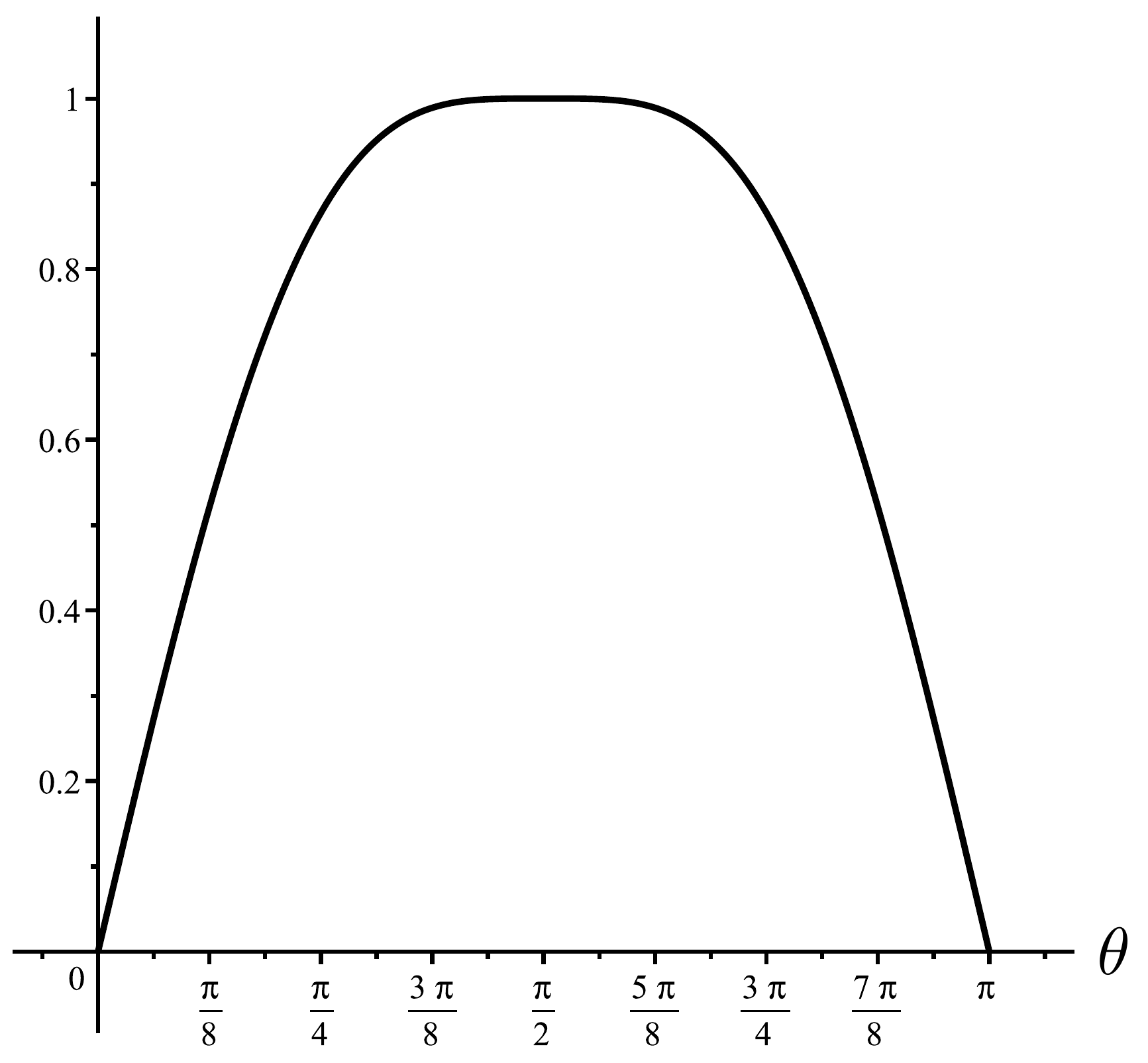}
\caption{Angular dependence of $g_{eff}$ to the non-polarized GW is depicted. 
Here, $\theta$ is the angle between the GW and the external magnetic field.}
\label{angle}
\end{figure}

The coupling constant $g_{eff}$ depends on the geometrical factor $\theta$, which is the angle between the magnetic field and 
the GW.
For example, when there is no polarization, i.e., $Q=U=V=0$, we have $g_{eff} \propto \sin\theta \sqrt{1+\cos^{2}\theta}$.
The angular dependence of $g_{eff}$ is depicted in Fig.\,\ref{angle}.
It has a maximum at $\pi/2$ where the GW propagates perpendicular to the external magnetic field.
It has the minimum value zero at $0$ and $\pi$, because then the GW propagates parallel to the external magnetic field and
cannot excite the spin precessions of electrons in such a configuration.
When the GW has polarization, i.e., $Q,\ U,\ V$ are nonzero, different angular dependence appears.

From the second term in Eq.\,(\ref{dri}), we see that 
GWs can induce the resonant spin precessions, that is,
excitations of magnons,  if the angular frequency of the GW is near the 
Lamor frequency  $2\mu_{B} B_{z}$.
This graviton-magnon resonance enables us to search for high frequency GWs in the GHz range~\cite{Ito:2019wcb,Ito:2020wxi}.
In the next section, we will estimate the sensitivity of the magnon GW detector.
\section{Sensitivity of magnon GW detectors} \label{sensitivity}
In this section, we evaluate the sensitivity of the magnon system to GWs.
To this end, we first need to take into account the dissipation effect in the system.
As the main dissipation effect to the magnons excited by GWs, 
we consider the thermal bath, namely, the photons in vacuum (cavity).
Such a dissipative effect can be taken into account 
by tracing out the thermal bath. 
Consequently, we obtain the master equation for the density matrix~\cite{walls}:
\begin{equation}
  \frac{d\rho}{dt} = -i \left[ g_{eff} \left(  \hat{c}^{\dagger} e^{i( \omega_{m} - \omega_{h})t} 
                                            +  \hat{c} e^{i(\omega_{h} - \omega_{m})t} \right) , \  \rho \right]
      + \frac{\gamma}{2} \left( 2\hat{c}\rho \hat{c}^{\dagger} - \hat{c}^{\dagger}\hat{c} \rho - \rho \hat{c}^{\dagger} \hat{c} \right) \ ,
      \label{master}
\end{equation}
where $\rho$ is the density matrix of the system, $\gamma_{m}$ is the line width of the magnon spectrum, 
and we defined $\omega_{m} = 2\mu_{B} B_{z}$.
Note that the interaction picture is used in the above equation.
Using the master equation (\ref{master}), we can calculate the average of the magnon number as
\begin{equation}
  <\hat{c}^{\dagger} \hat{c}> = {\rm tr}\left( \rho \hat{c}^{\dagger} \hat{c} \right)
                              = \frac{g_{eff}^{2}}{(\omega_{h} - \omega_{m})^{2} + \left(\gamma_{m}/2 \right)^{2}}
\end{equation}
Therefore, at the resonance $\omega_{h} = \omega_{m}$, the power of the excited magnons is given by
\begin{equation}
  P_{m} = \frac{4 \omega_{m}^{2} g_{eff}^{2}}{\gamma_{m}^{2}} .  \label{magpow}
\end{equation}

One can detect GWs by observing the excitations of magnons.
The simplest way to observe the excited magnons would be by measuring the resonance fluorescence of magnons, that is 
photons emitted from the magnons when they return to the ground state from the excited state.
Such a simple method is adopted in the context of the axion dark matter search~\cite{QUAX:2020adt,Flower:2018qgb}.
Another interesting method for observing the excited magnons is a quantum nondemolition detection of the magnons with 
a qubit~\cite{Lachance-Quirione1603150}.
Interestingly, quanta of magnons were detected with the quantum nondemolition technique in~\cite{Lachance-Quirione1603150} and 
it was also utilized for the axion dark matter search~\cite{Ikeda:2021mlv}.
To evaluate the ability of the magnon GW detectors, we consider the former method for the moment.
Let us prepare a cavity and put a ferromagnetic crystal inside.
The uniform magnon mode couples with a cavity mode.
Then, the power of the excited magnons (\ref{magpow}) converts into that of the cavity mode.
Here, we simply parameterize the conversion efficiency by $\eta$.
Although the original resonance frequency $\omega_{m}$ is slightly shifted in the magnon-cavity hybrid system,
we ignore it for simplicity.
From the input-output formalism~\cite{walls}, the output power of photons of the cavity mode which we actually observe is
\begin{equation}
  P_{{\rm out}} =  \frac{4 \eta Q_{m}^{2}  g_{eff}^{2}}{Q_{e}} \ ,
\end{equation}
where $Q_{e}$ represents the quality factor of the output load of the cavity, and
the magnon quality factor $Q_{m} = \omega_{m} / \gamma_{m}$ was defined.%
\footnote{More precisely, this quality factor should be replaced by that for the magnon-cavity hybridized mode. 
However, the order of the two quality factors would be the same because the quality factor of the magnon is worse in general.}

On the other hand, the system also has noise. 
The noise power can be characterized by the (effective) temperature by the Johnson-Nyquist relation:
\begin{equation} 
  N = b k_{{\rm B}} T_{N} \ ,  \label{john}
\end{equation}
where $b$ is a bandwidth.
In an observation with the observation time $\tau$, 
the noise power in the system is reduced to
\begin{equation}
  \sigma_{N} =   \frac{N}{\sqrt{b\tau}} = k_{{\rm B}} T_{N} \sqrt{\frac{b}{\tau}} \ ,  \label{radio}
\end{equation}
because we effectively have $b\tau$ samples in the observation.
In the case of the GW search, $b$ is given by 
$b = \frac{\omega_{m}} {2\pi \, {\rm max}(Q_{h}, Q_{i})}$.
We defined the quality factor of the GW $Q_{h}$ determined by
its coherence time: $\frac{2\pi}{\omega_{h}} Q_{h}$.
We also introduced $Q_{i}$ representing any quality factor of the apparatus such as the quality factors of the magnon and the cavity.
Using $\sigma_{N}$, the signal to noise ratio (SNR) is defined by
\begin{equation}
  {\rm SNR} = \frac{P_{{\rm out}}}{\sigma_{N}} 
            = \frac{4\sqrt{2\pi} \ \eta \ Q_{m}^{2}\  g_{eff}^{2} \, \sqrt{\tau}\ 
               {\rm max}(\sqrt{Q_{h}}, \sqrt{Q_{i}})}{k_{{\rm B}} T_{N} \sqrt{\omega_{m}} Q_{e}}\ .
\end{equation}
Let us now estimate the sensitivity of magnon detectors to GWs.
Since the coherence time of the GW depends on the source, we consider the most optimistic case where
the coherence time is comparable to or longer than the observation time. 
Then, we take the coherence time of the GW to be the observation time $\tau$.
As the ferromagnetic crystal, 
we take yttrium iron garnet, whose number density of electronic spin is $\sim 2.1 \times 10^{22} \, {\rm cm}^{-3}$.
We also assume no polarization of the GW, i.e. $h_{c} = h^{(+)} = h^{(\times)}$, and take the average of $\theta$.
By using the freedom to tune the magnitude of the magnetic field, 
we adopt the configuration which maximizes the form factor $C(\epsilon)$.
Then, setting ${\rm SNR} = 1$, the sensitivity of the magnon system to the amplitude of the GW reads%
%
%
\begin{eqnarray}
  h_{c} =  2.5 \times 10^{-20} 
    &\times& \left(\frac{T_{N}}{1 \, {\rm K}}  \right)^{1/2} 
             \left(\frac{V_{m}}{\left(4\pi/3\right) 10^{3} \, {\rm cm}^{3}}  \right)^{-1/2}
             \left(\frac{\eta}{1/2}  \right)^{-1/2}
             \left(\frac{C}{0.029}  \right)^{-1}  \nonumber \\
    &\times& 
            \left(\frac{\omega_{h}/2\pi}{1.6\times 10^{9} \, {\rm Hz}}  \right)^{-1}
            \left(\frac{Q_{e}}{10^{5}}  \right)^{1/2}
            \left(\frac{Q_{m}}{10^{4}}  \right)^{-1}
            \left(\frac{\tau}{1\, {\rm day}}  \right)^{-1/2},   \label{gcon}
\end{eqnarray}
where $V_{m}$ is the volume of the ferromagnetic crystal.
As one can see, for instance, in~\cite{Aggarwal:2020olq,Berlin:2021txa}, there are several 
high frequency GW detectors around the above frequency range $\sim$ GHz.
Our magnon GW detectors can be complementary to the detectors which use photons, as mentioned in the introduction,
to distinguish the observational signals from the axion dark matter and GWs.
Also, the magnon GW detector has different directional and polarization dependence of sensitivity (see Eq.\,(\ref{effco2}))
compared with the photonic GW detectors~\cite{Aggarwal:2020olq,Berlin:2021txa}.
Therefore, combining both the photonic and magnonic GW detector would be useful for identifying the direction and polarization 
of GWs.
Since the sensitivity (\ref{gcon}) is still below the predictions from the GW sources mentioned in 
the introduction, 
further efforts and strategies are needed to improve the sensitivity to reach theoretical predictions.
However, seeking high frequency GWs with the magnon GW detector with the above-noted sensitivity would still be interesting 
for exploring new physics.
For example, GHz GWs from extra dimensions~\cite{Seahra:2004fg,Clarkson:2006pq} may be detectable with the above sensitivity.
\section{Conclusion}

Detecting high frequency GWs is important not only for GW astrophysics and cosmology but also for fundamental physics.
To this end, new strategies are needed. 
We proposed magnon GW detectors in previous papers
where we utilized existing data from axion dark matter searches to impose constraints on GWs in the GHz range. 
In this paper, we have improved the analysis by considering an infinite sum of terms
in the expansion of Fermi normal coordinates. 
We showed that the form factor is most enhanced when the wavelength of the GW is comparable to the size of 
the ferromagnet, by evaluating the spatial integral exactly.
As a consequence, we obtained sensitivity of around $h_c \sim 10^{-20}$, which is much better than before.

There is still an open question of how to realize the configuration maximizing the form factor $C(\epsilon)$
where the wavelength of the GW is comparable to the size of the ferromagnet in real experiments.
It should be carefully considered whether one can realize the configuration while preserving the overall quality of the detector. 
Nevertheless, admittedly, for future detection of high frequency GWs, further improvement in sensitivity is needed.
Therefore, it is worth pushing forward GW detectors using condensed matter systems and seeking
possible strategies for improving the sensitivity.
In particular, employing quantum sensing~\cite{Degen_2017} may offer an important strategy, as already demonstrated in the magnon 
experiments~\cite{Lachance-Quirione1603150,Ikeda:2021mlv}.
\begin{acknowledgments}
A.\,I.\ was supported by World Premier International Research Center Initiative (WPI), MEXT, Japan, and JSPS KAKENHI Grant Number JP21J00162, JP22K14034.
J.\ S. was in part supported by JSPS KAKENHI Grant Numbers JP17H02894, JP17K18778, JP20H01902, JP22H01220.
\end{acknowledgments}

\bibliography{ref.bib}

\end{document}